# Interface Design Beyond Epitaxy:
# Oxide Heterostructures Comprising Symmetry-forbidden Interfaces

Hongguang Wang[1,†], Varun Harbola[1,†], Yu-Jung Wu[1],
Peter A. van Aken[1], Jochen Mannhart[1]*

[1]Max Planck Institute for Solid State Research,
Heisenbergstrasse 1, 70569 Stuttgart

Epitaxial growth of thin-film heterostructures is generally considered the most successful procedure to obtain interfaces of excellent structural and electronic quality between three-dimensional materials. However, these interfaces can only join material systems with crystal lattices of matching symmetries and lattice constants.

We present a novel category of interfaces, the fabrication of which is membrane-based and does not require epitaxial growth. These interfaces therefore overcome limitations imposed by epitaxy. Leveraging the additional degrees of freedom gained, we demonstrate atomically clean interfaces between three-fold symmetric sapphire and four-fold symmetric $SrTiO_3$. Atomic-resolution imaging reveals structurally well-defined interfaces with a novel moiré-type reconstruction.

[†] These authors contributed equally to this work
*Corresponding author: J.Mannhart@fkf.mpg.de

**Introduction**

Interfaces are used with great success to generate emergent electronic behavior, thus yielding phenomena that may differ qualitatively from those of the two abutting materials. Interface-engineered devices such as light-emitting diodes, photovoltaic solar cells, and field-effect transistors have revolutionized modern technology. Atomic-scale engineering of interfaces is therefore an important research topic (*1-4*). This research has already enabled the discoveries of quantum Hall effects (*5, 6*), giant magnetoresistance (*7*), and interfacial superconductivity (*8*), to name but a few examples.

Epitaxial growth of heterostructures has historically been the cornerstone for fabricating high-quality interfaces in three-dimensional (3D) materials. Unfortunately, this method imposes structural constraints by restricting the combinations of interfacing materials to those with compatible symmetries and lattice parameters (*9-11*). These constraints are so common that they are usually taken as an unquestionable given. However, mechanical stacking, a non-epitaxial growth method, is successfully used to create superb interfaces for two-dimensional (2D) materials. This stacking exploits that the van-der-Waals bonds between the 2D layers are weak. The stacking of 2D sheets has yielded groundbreaking discoveries (*12, 13*), because it allows virtually any twist angle between sheets, even between rather different materials (*14*). The corresponding advances have led to a plethora of discoveries, such as superconductivity, moiré excitons, interlayer magnetism, and moiré ferroelectricity generated at 2D interfaces (*15-21*). The success of this approach suggests to also search for new processes to overcome the epitaxial restrictions that apply to the fabrication of interfaces between 3D materials. In particular, it is intriguing to search for ways to join 3D materials with disparate lattice symmetries such that atomically sharp interfaces are obtained. We report on a way to achieve this goal by demonstrating atomically clean interfaces between $SrTiO_3$ and sapphire, interfaces that are epitaxially disallowed due to the underlying crystal lattice symmetries (cubic and hexagonal).

Our work builds on recent advances in fabricating membranes of 3D materials. Numerous research groups have fabricated and transferred single-crystalline membranes of 3D materials onto a variety of substrates (*22-34*). This progress has even produced interfaces that are sufficiently clean for electronic applications (*35*). Moreover, successful twisting and stacking of these membranes has been shown, where the overlaid lattices form moiré patterns (*36-39*).

However, current technologies are not entirely satisfactory for the fabrication of atomically clean and sharp interfaces. This limitation arises because the liquid processing methods during membrane fabrication expose the surfaces that will subsequently form the interfaces to solvents and environmental factors, resulting in contamination layers at the interfaces (*40*). Figure 1A shows a typical cross-sectional micrograph of such an interface prepared with state-of-the-art methods (see also Fig. S1). The interfacial contamination layer, which in this case is ~1.5 nm thick, is evident in the micrograph. This layer impedes the intended direct and uninterrupted contact between the membranes and their partner layers.

By extending the current technologies, we will show how atomically clean and structurally sharp interfaces can be fabricated between 3D materials, using the cubic-to-hexagonal $SrTiO_3$ – sapphire interface as a case study.

**Experiments and results**

To fabricate the targeted interfaces, Al-terminated step-and-terrace surfaces of (0001) sapphire crystals were prepared at $T = 1615\ °C$ by thermally annealing the crystals in ultrahigh vacuum (UHV) ($< 2 \times 10^{-8}$ mbar) with laser light ($\lambda$ ~10 μm) obtained from a $CO_2$ laser (see Materials and Methods in the Supplementary Materials and Fig. S2). Notably, the inherent absorbance of the infrared laser beam by sapphire eliminates the need to back-plate or coat the crystals.

Subsequently, [001]-oriented, 30-nm-thick $SrTiO_3$ membranes grown using pulsed-laser-deposition were manually transferred to the freshly prepared sapphire surface (see Materials and Methods in the Supplementary Materials and figs. S3, S4). The samples were then thermally annealed (1000 °C, 200 s, 0.1 mbar $O_2$), again by $CO_2$ laser irradiation in a UHV chamber, following precisely controlled temperature ramps (see Materials and Methods in the Supplementary Materials and figs. S5). The annealing temperature, which is set at 90% of the highest observed $SrTiO_3$ – sapphire dewetting temperature (*41, 42*), potentially allowed the interfaces to minimize their energy through cation intermixing or to induce an incommensurate structural interface reconstruction. Such a reconstruction would be compatible with an atomically clean interface structure. We fabricated three samples this way together with two standard samples for control and characterized these as discussed below and in the Supplementary Materials (Fig. S6, S7).

To analyze the chemical compositions of the interfaces and their microstructures, cross sections of the samples were viewed by high-annular dark-field scanning transmission electron microscopy (HAADF-STEM) imaging and characterized by electron energy loss spectroscopy (EELS).

Large-area STEM imaging clearly shows a pristine, straight, and sharp interface between the $SrTiO_3$ and the sapphire. Along the length of the sample measuring several micrometers in total, the interface is devoid of impurities or holes, see Fig. 1B and Fig. S8. The excellent uniformity of both sides of the interface fabricated as described and the stark chemical contrast at the interface shown by the EELS maps (Fig. S9) reveal—at the large scale imaged—a nominally perfect attachment of the $SrTiO_3$ to the sapphire without voids, interlayer contaminations, or buckling.

The inherent differences in the crystal structures of $SrTiO_3$ and sapphire pose challenges to atomically image both structures in a single projection. Therefore, atomic-resolution annular bright field (ABF) and HAADF-STEM imaging executed along $SrTiO_3$ [100], $Al_2O_3$ [1-100], $SrTiO_3$ [110] directions were taken to image all constituent elements simultaneously, see Figs. 2, 3A-B, and Fig. S10. These images reveal a ~3.5° twist angle between the two main axes of $SrTiO_3$ and sapphire. They show surprisingly well-ordered atomic lattices at the heterojunction and consistently exhibit an extremely clean interface.

The ABF images, particularly in the [100] orientation of $SrTiO_3$, reveal an intriguing feature, namely an apparent 2-Å-wide intensity minimum between $SrTiO_3$ and sapphire, see Figs. 3A-B. This dark band, reminiscent of a gap, adds a layer of complexity to our understanding of the interface, as discussed below.

The chemical composition obtained from EELS maps of the heterojunction taken along both orientations reveals a compositionally sharp interface between the $SrTiO_3$ and the sapphire within the experimental resolution, see Fig. 3C, and Fig. S11-S13. These maps confirm that the first cation layer of the $SrTiO_3$ at the interface is composed of Sr, whereas the uppermost cation layer of the sapphire top surface consists of Al. Note that there is no evidence of cation intermixing on either side of the interface. In contrast to the cation case, the EELS spectra suggest a reduced oxygen concentration in the first SrO layer of the $SrTiO_3$ at the interface (Fig. S14).

An intricate lattice distortion is observed at the $SrTiO_3$ – sapphire interface which reveals something quite remarkable. Perpendicular to the interface, the positions of all ionic planes match the expected values, except for the first two monolayers of the $SrTiO_3$. A STEM cross section of the bilayer projected along the sapphire [1-100] direction (Fig. 3B) exhibits an extraordinary alignment: the Sr columns within the $SrTiO_3$ layer extending in the interface plane align parallel to the Al columns of the sapphire observed in the region of the intensity minimum. This alignment persists despite a ~3.5° twist angle between the [100] direction of the $SrTiO_3$ membrane and the [1-100] direction of the sapphire. The pole figures shown in Figs. 4 and Fg. S15, taken on a sister sample with a 4° twist, capture this twist. In the SrO layer that bonds to the top Al layer of the sapphire, the Sr ions move toward the nearest Al ion to minimize the Coulomb energy; this relaxation induces a subtle wiggling within the Sr columns (Fig. S16). This clear atomic reconstruction of the $SrTiO_3$ lattice at the interface evidences ionic bonding between $SrTiO_3$ and sapphire.

In contrast, the Al columns in the sapphire are much less distorted, as shown in Fig. 3B, confirming the robust bonding within the sapphire crystals. The ionic rearrangement is therefore almost exclusively confined to the first SrO plane of the $SrTiO_3$, with the second SrO plane already being normally oriented nominally along the [001] axis of the $SrTiO_3$ bulk. The spatial arrangement of Sr and oxygen ions at the interface is dictated by the non-commensurate moiré pattern generated by the overlapping crystal lattices of the $SrTiO_3$ and sapphire with differing lattice constants and symmetries, see Fig. 5. Combined with the lack of mirror symmetry at the interface plane, the induced twist within the first unit cell of $SrTiO_3$ around the interface normal yields a chiral crystal structure of the first $SrTiO_3$ layer.

Furthermore, the HAADF and low-angle annular dark-field (LAADF) images reveal an enhanced LAADF signal in the first two $SrTiO_3$ unit cells, indicating local structural disorder or strain-induced electron beam dechanneling(*43-45*), see Fig. S17. Quantitative analyses reveal an out-of-plane tensile strain of 2.8% with unchanged in-plane lattice parameters, resulting in a significant but compensated polarization in the second $SrTiO_3$ monolayer (~50 $\pm$ 10 μC/cm$^2$), see figs. S17, S14. Although this localized reconstruction is analogous to the coherent strain induced at epitaxial interfaces, it is impossible to obtain by epitaxial growth.

Consistent with the clean and abruptly changing microstructure of the interface, the X-ray diffraction (XRD) pole figures are striking, as they embody both the three-fold symmetry of the sapphire substrate and the four-fold symmetry of the $SrTiO_3$ layer within a single diffraction pattern, see Fig. 4 and Fig. S15. Heterostructures with such pole figures are disallowed by epitaxial growth because such growth must maintain the inherent in-plane symmetry across the interface.

**Discussion and Outlook**

To fabricate atomically sharp interfaces that defy the lattice symmetry constraints of conventional epitaxial growth of 3D materials, we combined (a) epitaxial growth and lift-off techniques to obtain freestanding oxide membranes, (b) meticulous substrate preparation resulting in high-quality single-termi-nated surfaces, and (c) post-deposition annealing at high temperatures. $SrTiO_3$ – sapphire heterojunctions serve as our case study of interfaces between materials featuring cubic and hexagonal symmetries. Analyses of these interfaces using atomic-resolution STEM and EELS revealed their exceptional structural homogeneity, cleanness, and atomic sharpness. Surprisingly, the interfaces exhibit a novel structural configuration. The Sr ions of the Sr-O plane next to the interface shift towards their closest Al neighbors and thereby locally deform the moiré lattice generated by the juxtaposition of the square and hexagonal crystal lattices. This shift of the Sr ions therefore introduces a chirality into the $SrTiO_3$ layer at the interface. The quasi-crystalline interfacial reconstruction of the interfacial cations allows the lattice distortion to be incoherent on the bulk lattice scale, thus enabling the observed lattice relaxation within a single monolayer.

The preparation process presented here eliminates the stringent necessity of epitaxial growth that requires high-quality interfaces between 3D materials to be fabricated only between materials that have matching symmetries of their crystal lattices. Furthermore, the epitaxial constraint of fitting lattice constants becomes obsolete. Clearly, not all combinatorially possible material pairings are applicable for such interfaces because additional constraints associated with epitaxial growth such as chemical compatibility, matching thermal expansion, and lack of interdiffusion at fabrication temperatures remain in effect.

Whereas the moiré-type structural rearrangement of ionic positions at interfaces between ionic crystals reported here has been found by combining crystal lattices with three-fold and four-fold symmetries, we expect corresponding phenomena to occur also when crystals of other symmetries are joined, as long as strong bonding is established across the interfaces. Owing to mismatched symmetries, interfaces created in this way will yield relaxed thin films with saturated bonds on well-defined surfaces, rather than extended defects such as edge dislocations. The degrees of freedom to connect lattices of different lattice symmetries and lattice constants opens a very promising novel phase space to create innovative electron systems with outstanding properties at interfaces between 3D materials.

Looking ahead, the unequal symmetries of the materials abutting such interfaces promise novel electronic and magnetic behavior, such as new forms of 2D electron systems, multiferroicity or topological phenomena, and symmetry-induced frustration possibly affecting their properties. Moreover, the flexibility to select membrane – substrate twists introduces an entirely new realm of potential moiré patterns and related phenomena, see Fig. 5. In addition, novel transport phenomena such as current flow across Josephson junctions as well as distinct electronic behaviors within superlattices formed by stacking such bilayers are expected to emerge as exciting avenues for future exploration.

**Acknowledgments**

The authors thank Hans Boschker, Seungwon Jung, and Mathias Scheurer for insightful discussions, Julia Deuschle and Ute Salzberger for assistance with TEM sample preparation, Benjamin Stuhlhofer for the X-ray pole figure data, and L. Pavka for editorial support. This project was partially supported by EU FLAG-ERA Project To2Dox and the German Science Foundation (DFG).

50. A. Ohtomo, D. A. Muller, J. L. Grazul, H. Y. Hwang, Artificial charge-modulationin atomic-scale perovskite titanate superlattices. *Nature* **419**, 378-380 (2002).

51. K. Tomita, T. Miyata, W. Olovsson, T. Mizoguchi, Strong excitonic interactions in the oxygen K-edge of perovskite oxides. *Ultramicroscopy* **178**, 105-111 (2017).

52. J. Wei *et al.*, Direct Measurement of Electronic Band Structures at Oxide Grain Boundaries. *Nano Letters* **20**, 2530-2536 (2020).

53. M. Wu *et al.*, Engineering of atomic-scale flexoelectricity at grain boundaries. *Nature Communications* **13**, 216 (2022).

54. S. C. Abrahams, S. K. Kurtz, P. B. Jamieson, Atomic Displacement Relationship to Curie Temperature and Spontaneous Polarization in Displacive Ferroelectrics. *Physical Review* **172**, 551-553 (1968).

**Figure Legends**

**Fig. 1. Large-area cross sectional imaging of heterostructures.**

Large-area HAADF-STEM cross sectional images of fabricated SrTiO$_3$–sapphire heterojunctions. Sample A was prepared in the standard way and shows an ~1.5-nm-thick contamination layer (blue) at the interface. Sample B, prepared with the method presented here, is characterized by a clean interface between the SrTiO$_3$ and the sapphire. The images show raw data; the color scale renders the brightness directly. The arrows mark the nominal interface between the sapphire and the SrTiO$_3$.

**Fig. 2. Sharp heterointerface between SrTiO$_3$–sapphire.**

HAADF-STEM images of cross sections of SrTiO$_3$–sapphire heterojunctions viewed along (A) SrTiO3 [100] and (B) sapphire [1-100]. The arrows mark the nominal interface between the sapphire and the SrTiO$_3$.

**Fig. 3. Atomic structure and chemistry at the interface.**

Atomic-resolution STEM images of the SrTiO$_3$–sapphire interface viewed along the (A) SrTiO$_3$ [100] and (B) sapphire [1-100] directions. The left and right panels correspond to HAADF and inverted ABF images, respectively. The arrows at the very left mark the interface. The image or the right panel in A reveals a dark band, reminiscent of a gap of ~2 Å between the top Al layer of the sapphire and the bottom Sr layer of the SrTiO$_3$, as discussed in the text. The yellow arrow in (B) marks the position of a Sr column at the very bottom layer of the SrTiO$_3$. As the other Sr columns do, this column aligns parallel to the Al column at the top of the sapphire (green arrow). (C): Atomic-resolution STEM EELS maps of the constituent elements O, Al, Sr, Ti and their composite map taken for an interfacial cross section.

**Fig. 4. X-ray pole figure.**

X-ray pole figure taken on a SrTiO$_3$–sapphire sample with a ~4° twist between the main crystal lattices. The figure clearly shows reflection peaks simultaneously arising from the hexagonal crystal lattice of the sapphire and the square SrTiO$_3$ lattice. The same data are presented with further information in Fig. S15.

**Fig. 5. Stacking and twisting different symmetries.**

Moiré patterns caused by superpositions of two identical hexagonal lattices (left) and a hexagonal lattice with a square lattice (right). The twist angles between the main axes of the lattices is 3° (top) and 11° (bottom) in both cases. The areas colored yellow have been added to highlight the moiré structures. Whereas the hexagonal–hexagonal moiré lattices show the well-known repetitive six-fold pattern, the superposition of the hexagonal and square lattices induces quasi-periodic stripe-like patterns.

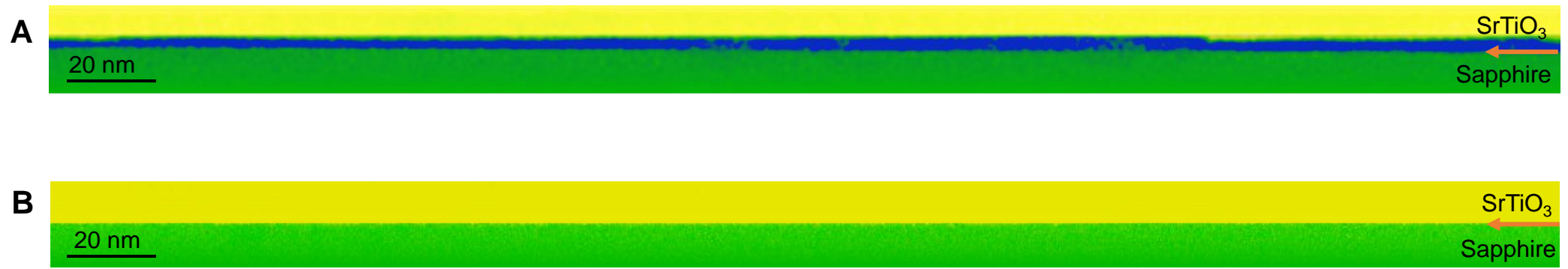

*Wang et al., Figure 1*

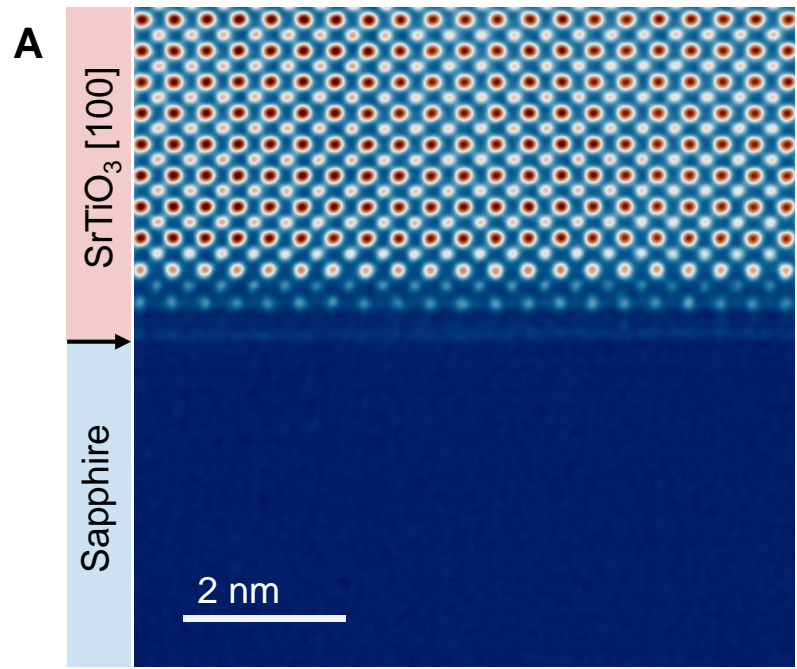 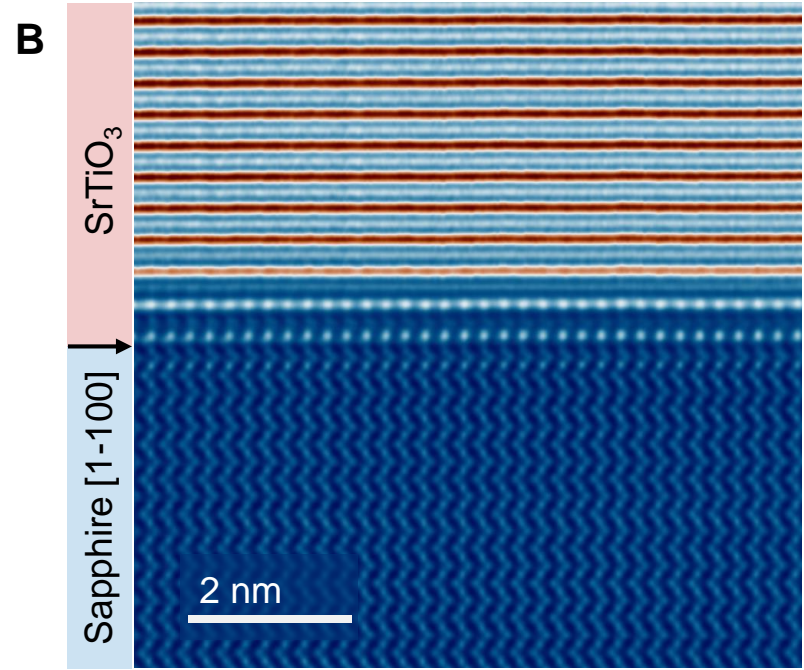

*Wang et al., Figure 2*

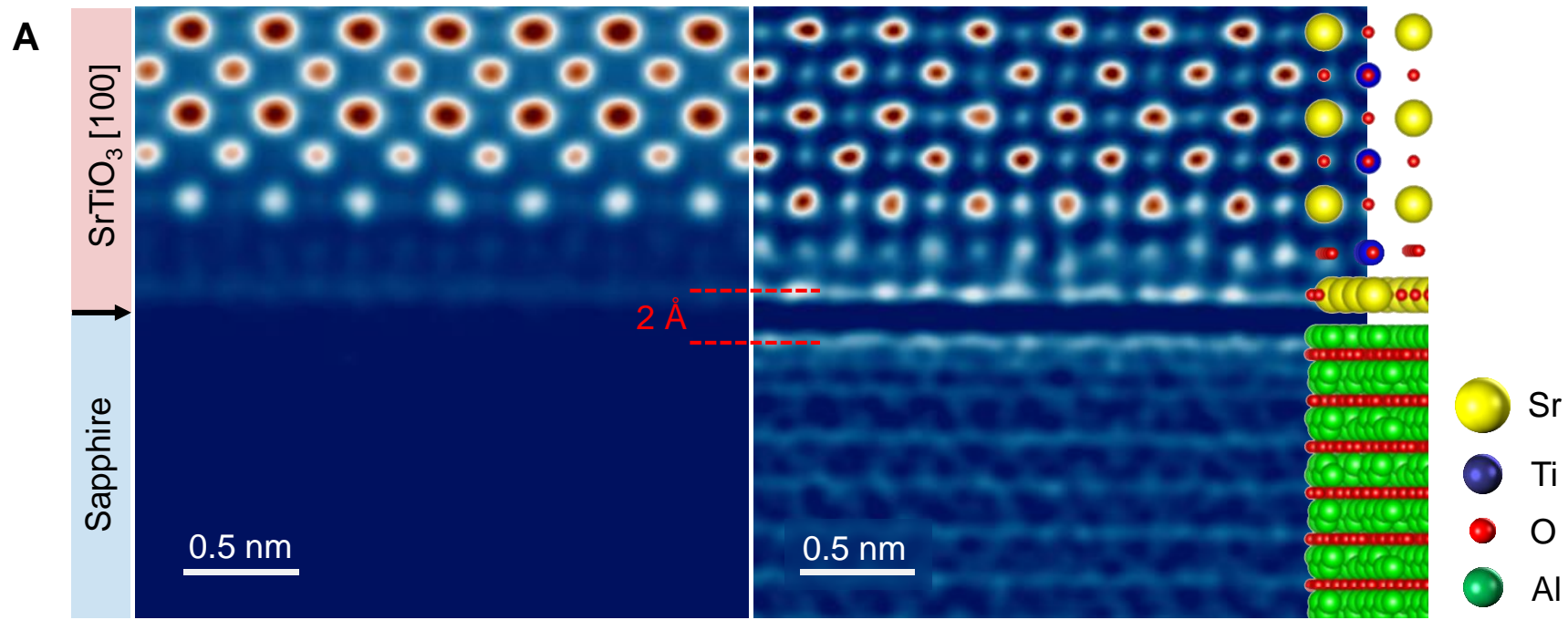

Wang et al., Fig. 3A

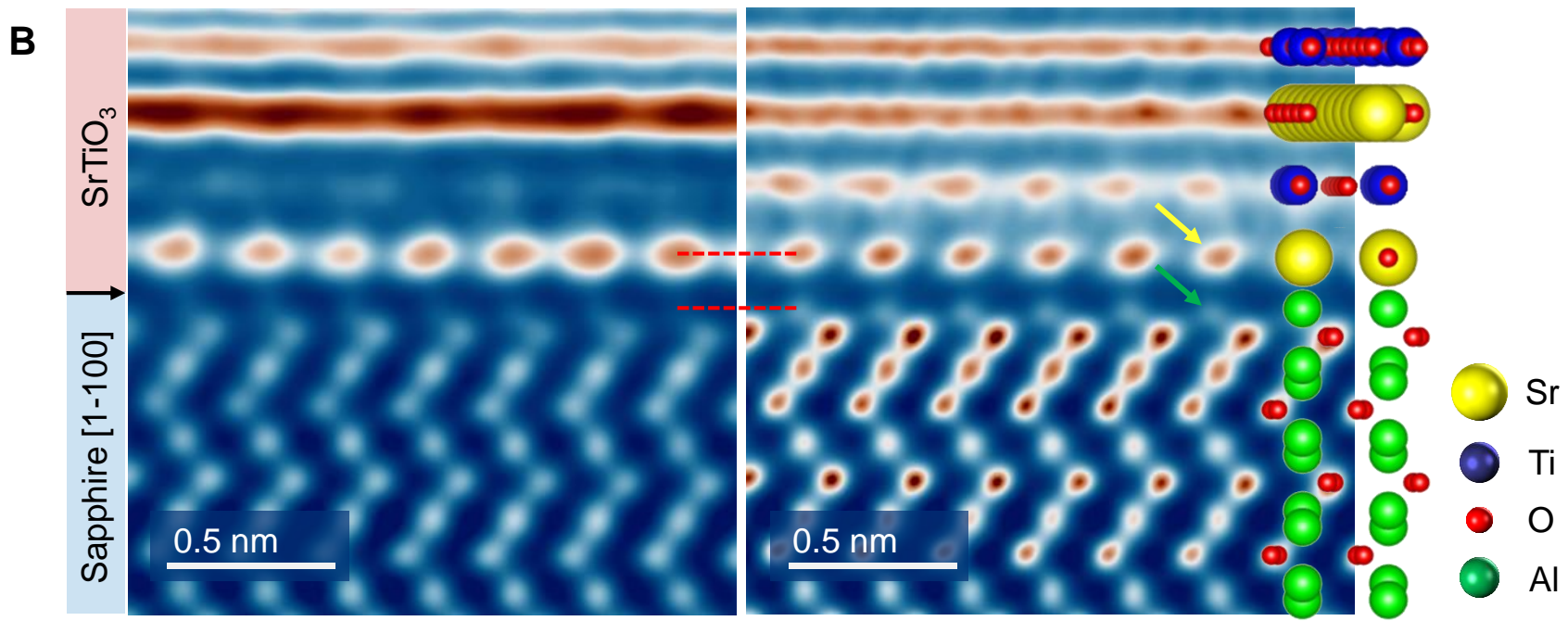

Wang et al., Fig. 3B

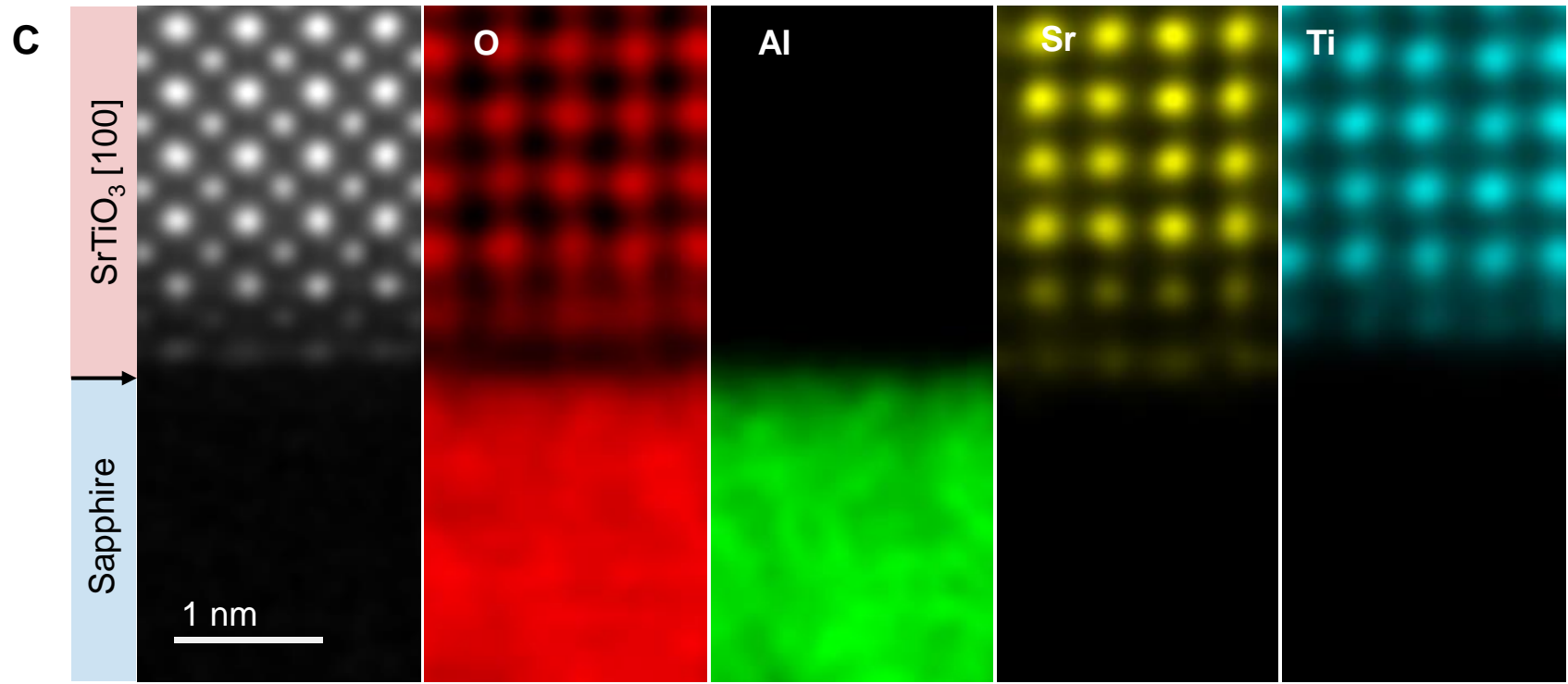

*Wang et al., Fig. 3C*

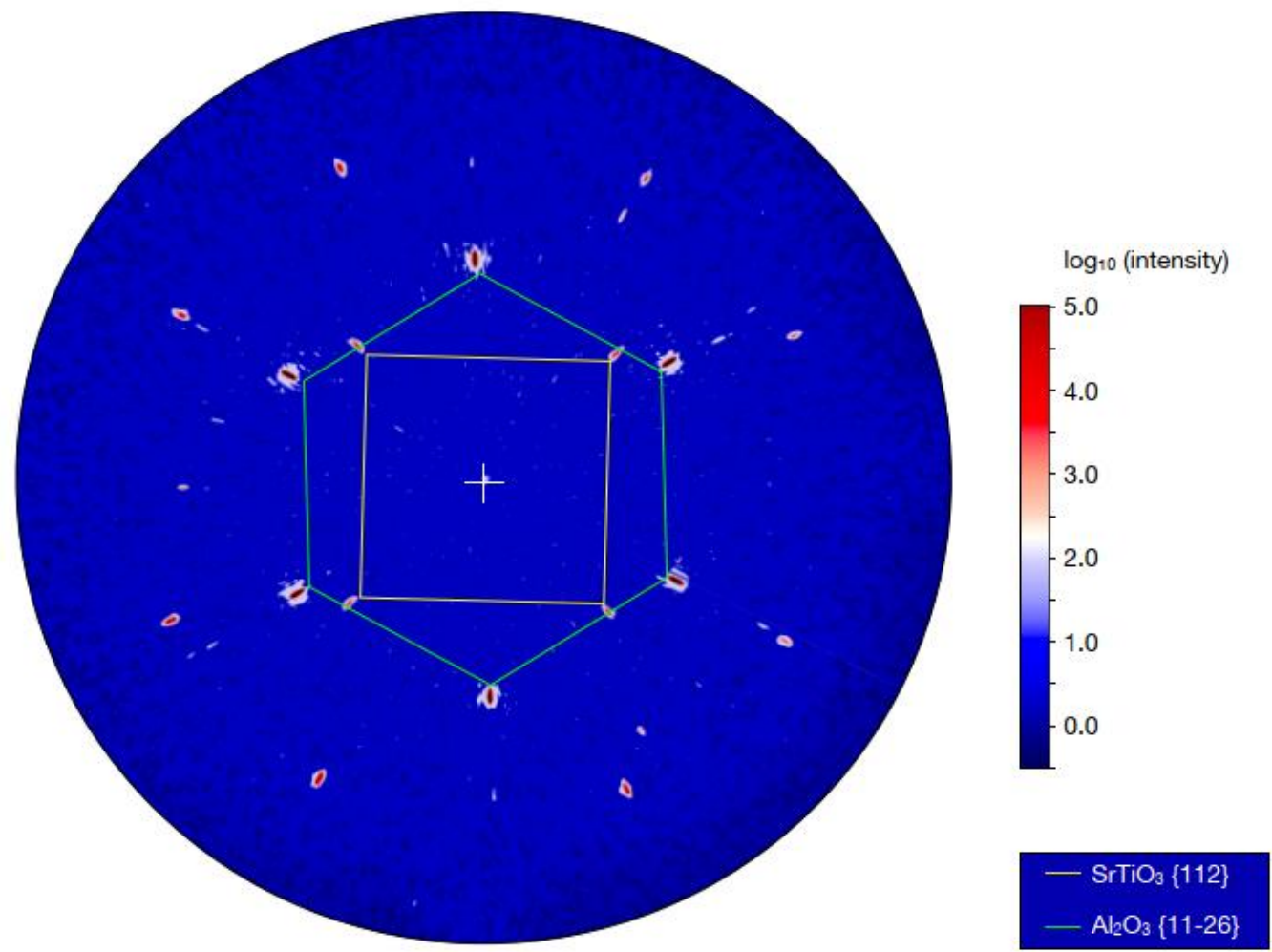

*Wang et al., Figure 4*

| | Honeycomb-Honeycomb | Square-Hexagon |
|---|---|---|
| 3° | 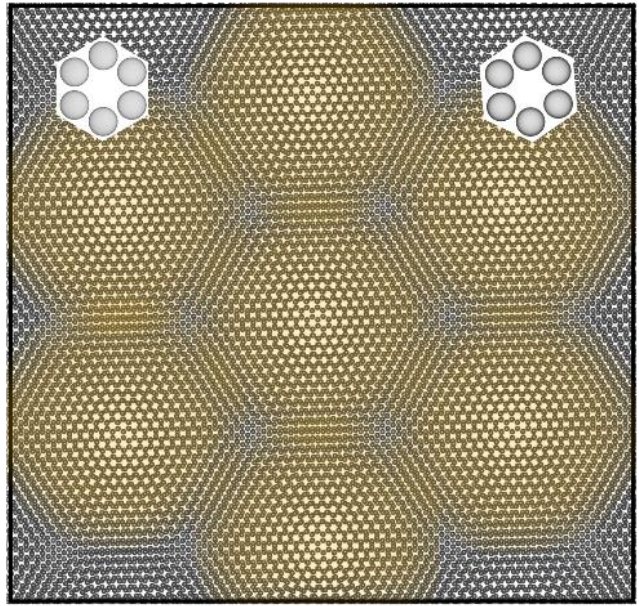 | 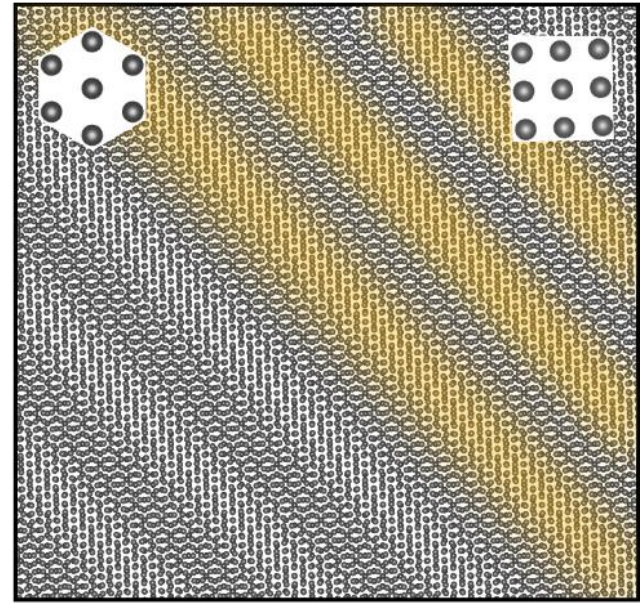 |
| 11° | 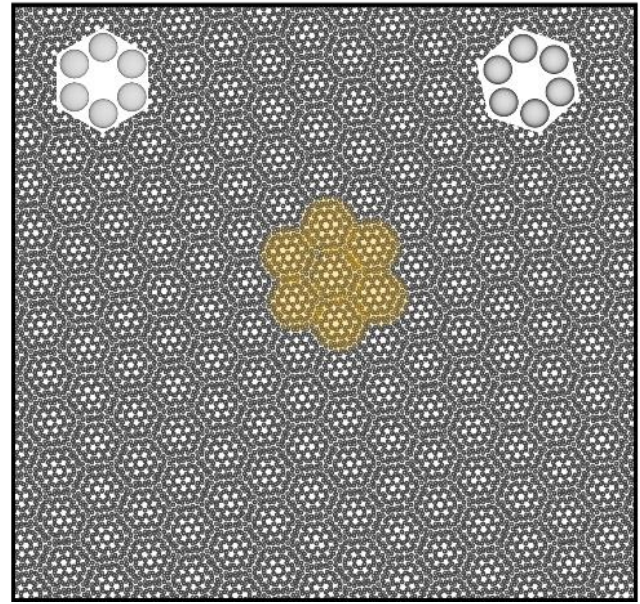 | 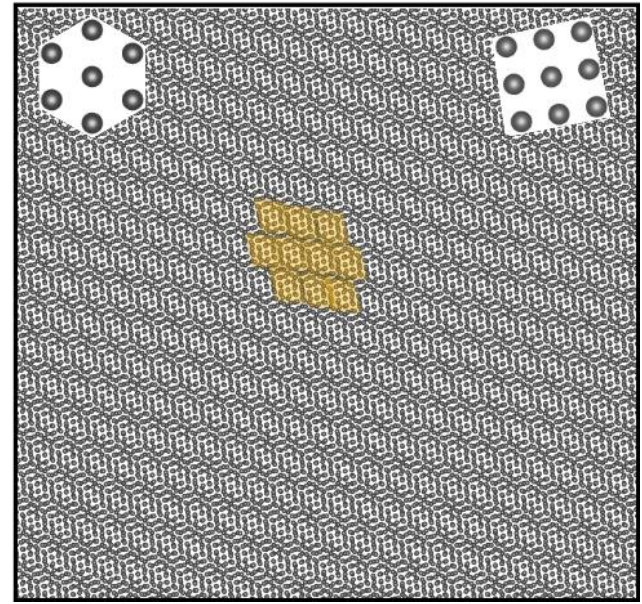 |

*Wang et al., Figure 5*

# Supplementary Materials for

## Interface Design Beyond Epitaxy:
## Oxide Heterostructures Comprising Symmetry-forbidden Interfaces

Hongguang Wang[1†], Varun Harbola[1†], Yu-Jung Wu[1],

Peter A. van Aken[1], Jochen Mannhart[1*]

[1]Max Planck Institute for Solid State Research,

Heisenbergstrasse 1, 70569 Stuttgart

*Corresponding author: J.Mannhart@fkf.mpg.de

[†] These authors contributed equally to this work

**The PDF file includes:**

Materials and Methods

Supplementary Notes

Figs. S1 to S20

References

**Materials and Methods**

Thin film growth

Epitaxial heterostructures consisting of $Sr_3Al_2O_6$ sacrificial layers and $SrTiO_3$ films were grown by pulsed laser deposition using a 248 nm laser. Before growth, the $SrTiO_3$ (001) substrate surfaces were prepared by annealing (direct $CO_2$ laser heating, 1200 °C, 480 s $O_2$ pressure: $7.5 \times 10^{-2}$ mbar) to achieve atomically flat single-terminated surfaces. The 8 nm thick sacrificial layer was grown on the thus annealed substrates at a substrate temperature of 825 °C, a $PO_2$ of $1 \times 10^{-5}$ mbar, and 1.8 J cm$^{-2}$ laser fluence, and a repetition rate of 1 Hz. All targets were obtained from the Kurt J. Lesker company. The films were grown in the layer-by-layer mode, their growth was monitored by in-situ RHEED (Fig. S3).

Thermal laser annealing

The sapphire substrate surfaces were prepared by annealing in UHV with direct $CO_2$ laser heating at 1615 °C to achieve a double-step termination of the surface before transferring the membrane onto the substrate. The annealing temperature was reached by ramping the substrate temperature from room temperature to 1400 °C at 4 K/s, then to 1615 °C at 1 K/s. After the 200 s anneal at 1615 °C the laser power was turned off completely, yielding an initial cooling rate of several hundred K/s. With this procedure, the sapphire surface acquires a step-and-terrace structure, with step heights equalling the distance between 2 Al layers along the sapphire c-axis (Fig. S2).

Membrane transfer and post anneal

PMMA layers were spin-coated onto the heterostructures comprising the $Sr_3Al_2O_6$ and $SrTiO_3$ films. The $Sr_3Al_2O_6$ layer was then dissolved by immersing these samples into deionized water held at room temperature. After lift-off with the PMMA support, the freestanding membranes were attached to the freshly annealed $Al_2O_3$ substrates by heating at 80 °C on a hot plate. The membranes remained on the substrates after dissolving the PMMA layers in acetone and a subsequent clean in isopropanol. These structures were transferred into the UHV chamber to post-anneal in 0.1 mbar $pO_2$ pressure for 200 s. The temperature for this post-anneal was 1000 °C except for one sample that was annealed at 1100 °C (Figs.

S6, S7).

Atomic force microscopy

The AFM characterization was done using an Asylum Cypher AFM in tapping mode and tapping-mode cantilevers with a nominal frequency of 300 kHz (Tap300 DLC, Budgetsensors).

X-ray diffraction

The X-ray diffraction data were acquired with a monochromated Cu-K$_{\alpha 1}$ source on a D8 Bruker system. The sample for taking the pole figure was mounted on a 4-axis gimbal for $X$ and $\Phi$ scans. The 2$\Theta$= 57.812° was set to capture both the SrTiO$_3$ {112} set of planes, and also the Al$_2$O$_3$ {11-26} set of planes in the same pole figure.

TEM sample preparation

TEM specimens of the fabricated heterostructures were prepared in two different ways: (a) by focused ion beam (FIB) processing or (b) by mechanical wedge polishing followed by Ar ion-beam milling at L-N$_2$ temperature.

The FIB preparation (a) was performed using a FEI scios dual beam system equipped with a Ga source (Fig. S18). Prior to the FIB cutting, the top surface was coated with carbon for protection. The milling steps were performed with a 30 kV Ga beam, for thinning a voltage of 5 kV was used. The final polishing step was done with a 2 kV Ga beam to obtain a TEM lamellae with the thickness of the region of interest below 30 nm.

For procedure (b), the sample was first cut into slabs of 0.5 mm width and 1.5 mm length with a wire saw (Well, Model 3242). A thin layer of glass was then glued to the surface of the slab (Epoxy bond 110) for protection. Next, an Allied MultiPrep System was employed for automated tripod polishing of the samples in cross-sectional geometry using diamond lapping films with grain sizes of 3.0, 1.0, 0.5, and 0.1 µm. Afterward, a Gatan Precision Ion Polishing System (PIPS II, Model 695) was used for Ar$^+$ ion milling of the samples at L-N$_2$ temperature. To thin the sample and to reduce the ion-beam-induced damage, the acceleration voltage during Ar$^+$ ion milling was progressively lowered from 3.3 kV to 2.0 kV and 0.3 kV. Optical microscopy images (Zeiss AxioCam HRc) were used to image the thin region of the final TEM specimen. Optical interference fringes indicated that the thickness close to the curved region is thin enough for TEM observation (Fig. S19).

Scanning transmission electron microscopy

TEM investigations were carried out using a spherical aberration-corrected STEM (JEM-ARM200F, JEOL Co. Ltd.) equipped with a cold field emission gun and a DCOR probe Cs-corrector (CEOS GmbH) operated at 200 kV. The images were obtained using JEOL ADF and BF detectors with a probe size of 0.8 Å (corresponding to the spot size 8C in the experimental setting), a convergent semiangle of 20.4 mrad, and a camera length of 6 cm. The corresponding collection semiangles for HAADF, LAADF, and ABF imaging were 70–300 mrad, 40-100 mrad and 11–22 mrad, respectively. To improve the signal-to-noise ratio (SNR) and to minimize the image distortion of HAADF and ABF images, 6 serial frames were acquired with a short dwell time (2 µs/pixel), aligned, and added afterwards. The obtained STEM images were subsequently denoised using a bandpass filter. Atomic column positions were determined with 2D Gaussian fitting, allowing for the measurements of the IP and OOP lattice parameters and atomic displacement with high accuracy (Fig. S17).

Electron energy-loss spectroscopy

EELS acquisition was performed with a Gatan GIF Quantum ERS imaging filter equipped with a Gatan K2 Summit camera and a CCD camera with a convergent semiangle of 20.4 mrad and a probe size of 1 Å (5C). With the 1.5 cm camera length and 5 mm entrance aperture of the image filter, a collection semiangle of 111 mrad was used for EELS acquisition. EELS spectrum imaging was performed with a dispersion of 0.5 eV/channel for the simultaneous acquisition of signals of the Ti-L$_{2,3}$, O-K, Al-K, and Sr-K edges. The raw-spectrum image data were denoised by applying a principal component analysis (PCA) with the multivariate statistical analysis (MSA) plugin (HREM Research Inc.). Dual-EELS acquisition was conducted using a CCD camera with a dispersion of 0.1 eV/channel to acquire EELS spectra of Al-L$_{2,3}$, O-K, and Ti-L$_{2,3}$ edges for further fine structure analysis. The electron energy-loss near-edge structure analysis of Ti-L$_{2,3}$ was performed using a multiple linear least-squares fitting algorithm. For calculating the Ti valence state, the electron energy-loss near-edge structure of Ti-L$_{2,3}$ was analyzed using a multiple linear least-squares fitting algorithm based on the measured Ti-L$_{2,3}$ edges of standard samples with Ti$^{4+}$ (SrTiO$_3$) and Ti$^{3+}$ (LiTiO$_2$) (Fig. S20).

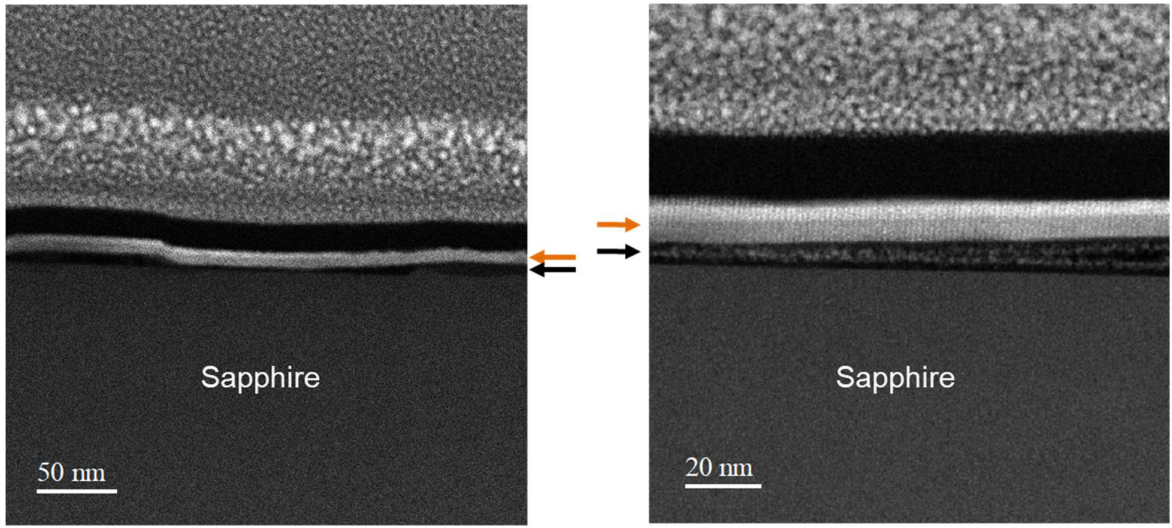

**Fig. S1** ADF-STEM image of the cross-section of a transferred oxide membrane/sapphire heterojunction. This membrane marked by the range arrow consists of SrTiO$_3$ capped by La$_{0.67}$Sr$_{0.33}$MnO$_3$. During the transfer process, all surfaces forming the interface were subjected to chemicals and air so that the interface is typically contaminated by impurities such as hydrocarbons and water. These contamination layers (as indicated by black arrows) may even cause buckling of the transferred membranes.

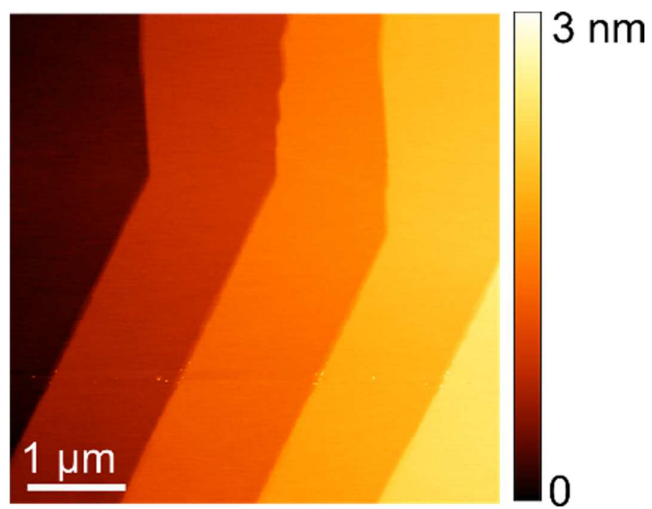

**Fig. S2.** AFM image of the surface of a sapphire substrate taken after annealing the substrate at 1615 °C for 200s. The surface shows sharp step edges and atomically smooth terraces. The RMS roughness is ~131 pm.

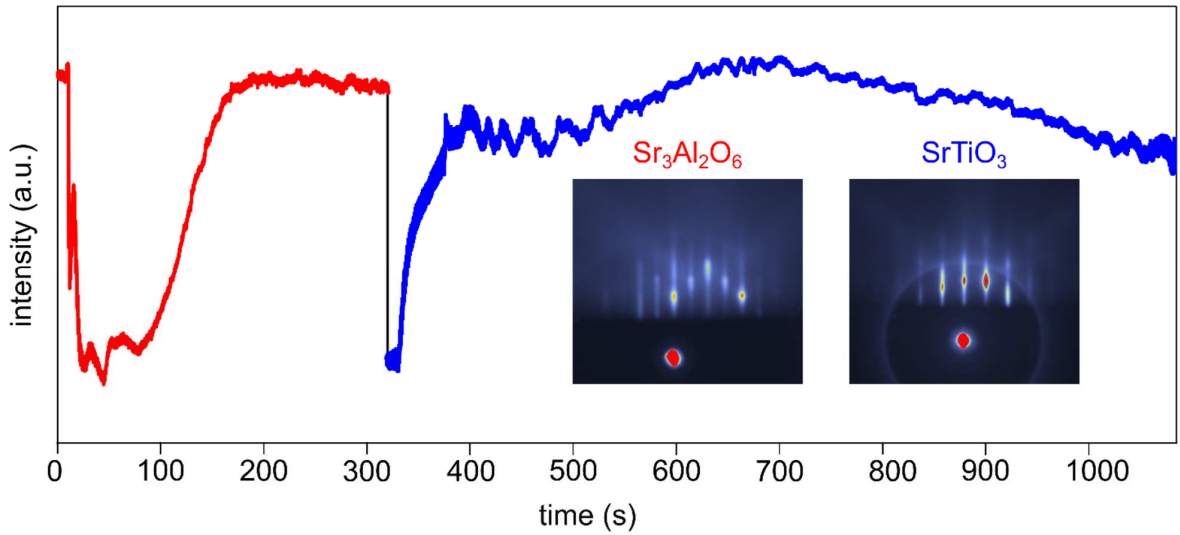

**Fig. S3**. RHEED intensity oscillations measured during the growth of an 8-nm-thick $Sr_3Al_2O_6$ sacrificial layer (red) and a 30-nm-thick $SrTiO_3$ membrane film (blue) on $SrTiO_3$ substrate. The RHEED oscillations reveal layer-by-layer growth. Inset: RHEED patterns taken after deposition of the $Sr_3Al_2O_6$ layer (left) and $SrTiO_3$(right).

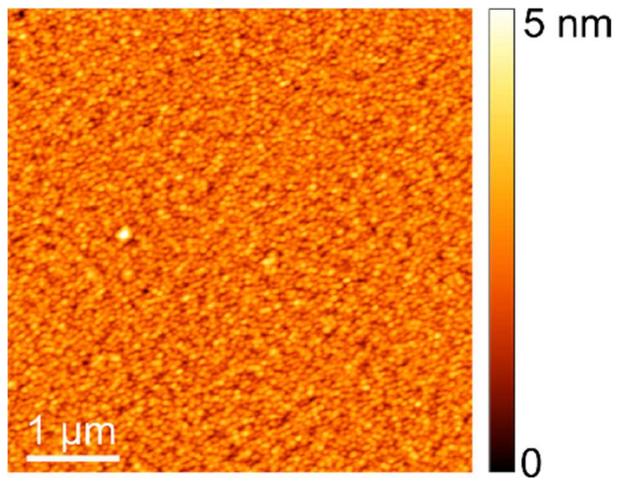

**Fig. S4.** AFM image of the surface of a film on a sacrificial layer bilayer (30-nm-thick $SrTiO_3$ on 8-nm-thick $Sr_3Al_2O_6$ deposited on a $SrTiO_3$ substrate. The RMS roughness is ~466 pm.

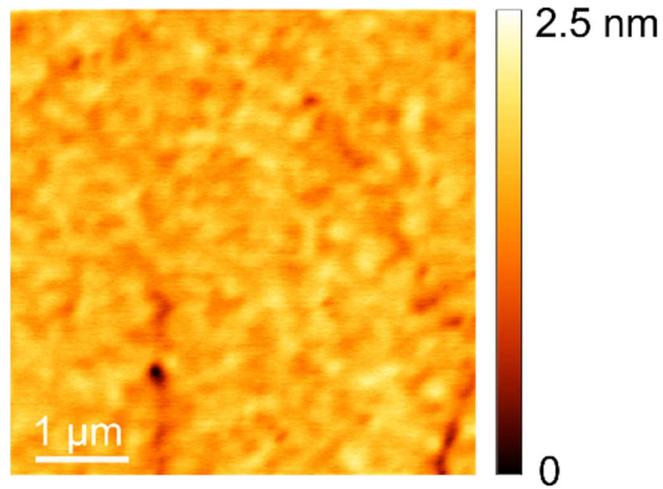

**Fig. S5.** AFM image of a SrTiO$_3$ film transferred onto a c-cut sapphire substrate. Before taking the image the sample was annealed at 1000 °C for 200 s using pO$_2$= 0.1 mbar. The RMS roughness is ~200 pm. This is the sample that is discussed in most detail in the main text.

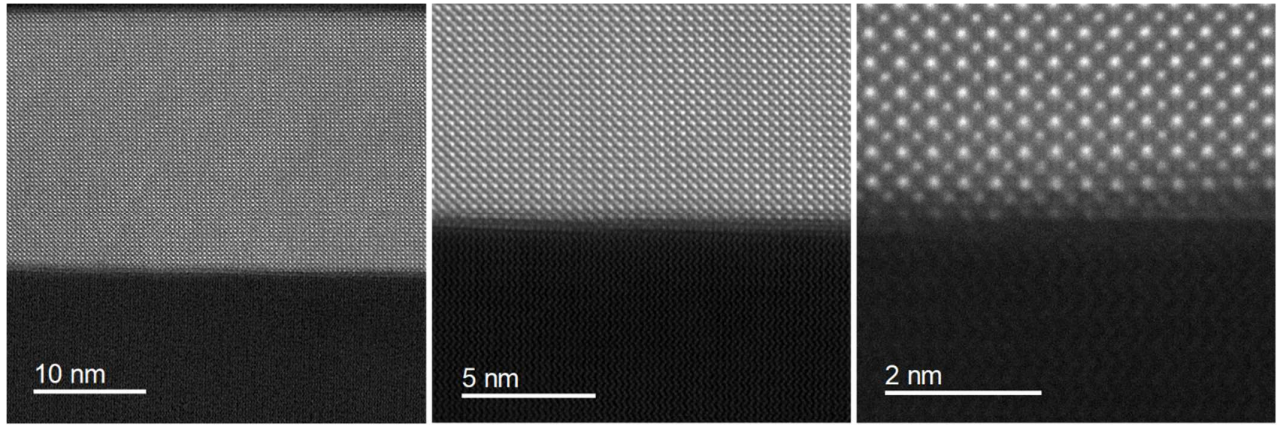

**Fig. S6.** ADF-STEM images of the cross section of another SrTiO$_3$/sapphire heterojunction annealed at 1000 °C. The micrographs reveal a straight and sharp interface.

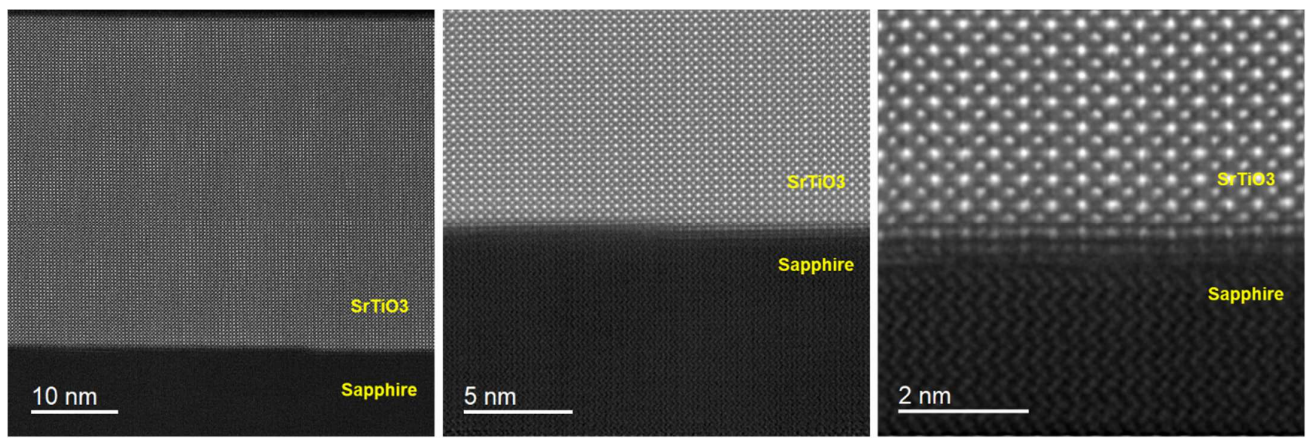

**Fig. S7.** ADF-STEM images of the cross section of a SrTiO$_3$/sapphire heterojunction annealed at 1100 °C. The micrographs reveal a straight and sharp interface.

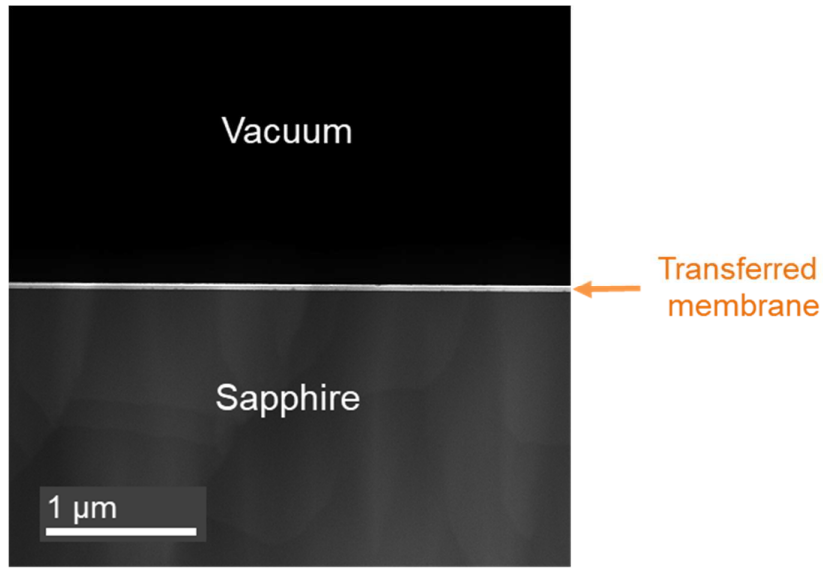

**Fig. S8.** ADF-STEM image of the cross section of the sample discussed in the main text. The image shows that the membrane is flat and forms a straight interface with the sapphire over a lateral length scale of at least 3 $\mu$m.

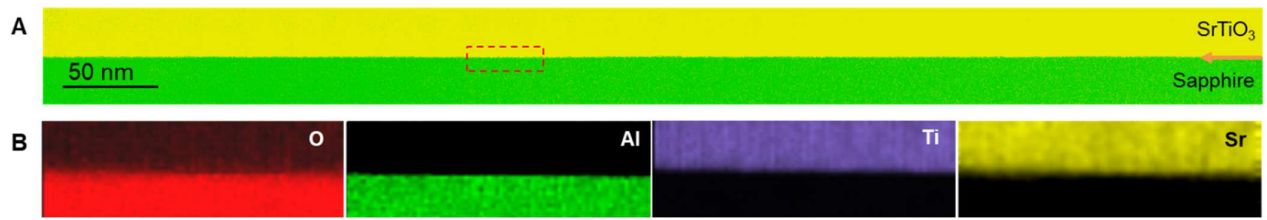

**Fig. S9.** (A) Low-magnification HAADF-STEM image of the cross section of the sample discussed in the main text. (B) STEM-EELS maps of the constituent elements O, Al, Ti, Sr, measured in the region marked by the red rectangle in (A).

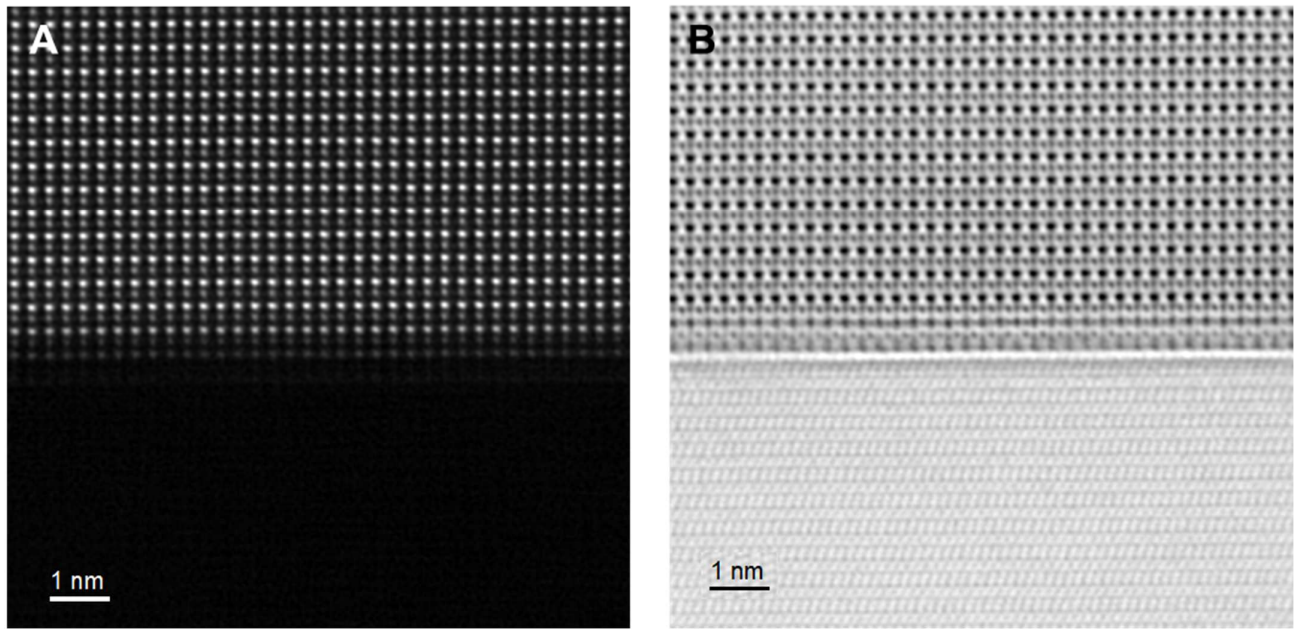

**Fig. S10**. Atomic-resolution HAADF-STEM images of the interface region of the sample discussed in the main text viewed along the SrTiO$_3$ [110] directions. As discussed in the main text, the sample shows a reconstruction of the first two SrTiO$_3$ layers next to the interface.

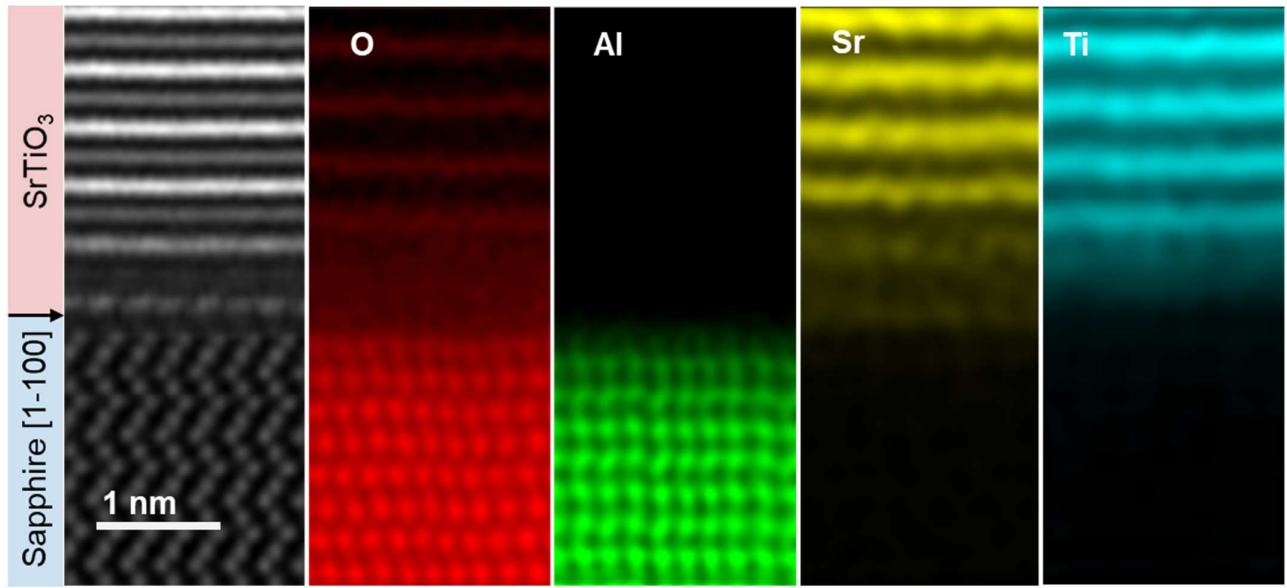

**Fig. S11.** Atomic-resolution STEM-EELS maps of the cross section of the sample discussed in the main text, presenting the constituent elements O, Al, Sr, and Ti along the Al$_2$O$_3$ [1-100] direction.

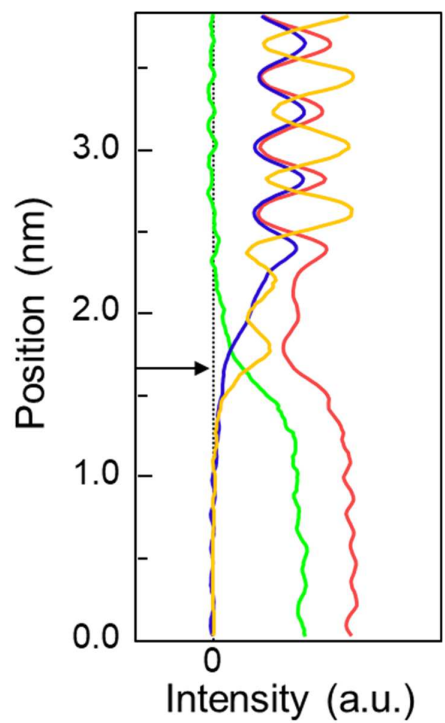

**Fig. S12.** EELS signal profiles taken at a sample cross section covering the interface. The colored curves present the EELS signals of O (red), Al (green), Sr (yellow), and Ti (blue). The black horizontal arrow marks the interface.

**Supplementary Note 1:** Weak spreading of the elemental signals occurs within 0.4 nm around the interface. This is also the length scale of the signal broadening due to multiple electron scattering in this 30-nm-thick sample (Fig. S13) (*46-48*). The data shown therefore reveal that the interface is atomically sharp, with no observable elemental intermixing occurring on either side of the interface.

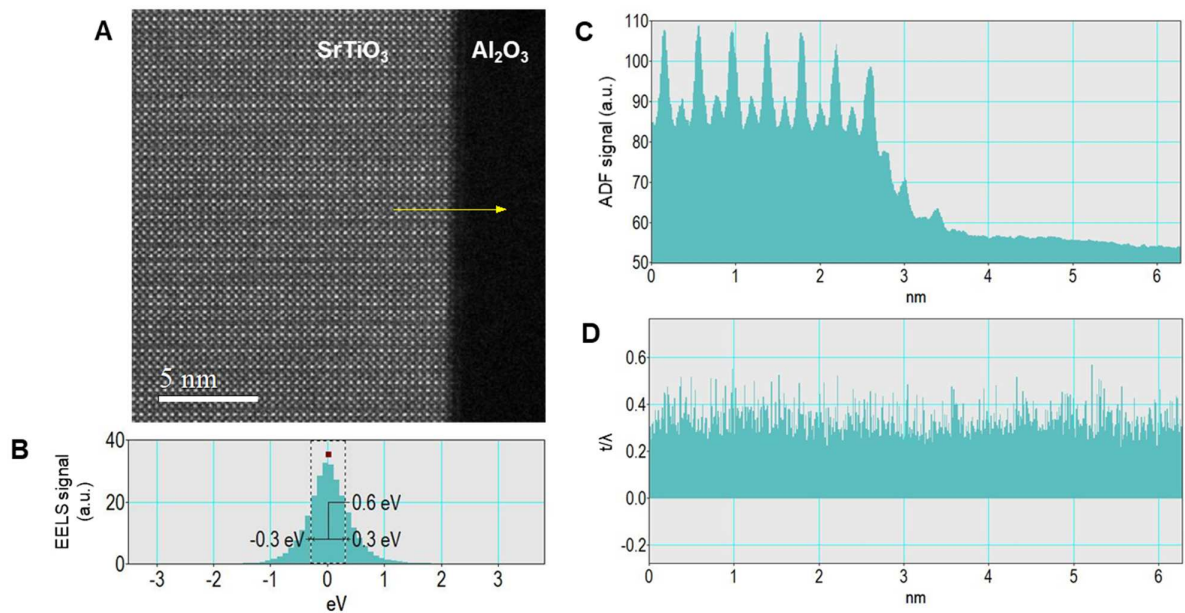

**Fig. S13.** EELS studies and thickness measurement taken at a cross section of a SrTiO$_3$/sapphire heterojunction. (A) HAADF-STEM images around the interface between SrTiO$_3$ and sapphire. The yellow arrow marks the direction of the EELS line scan measurement. (B) The zero-loss peak of the EELS data at the dispersion of 0.1 eV/channel, demonstrating an energy resolution of 0.6 eV. (C) Profile of the ADF signal intensity taken along the line scan direction. (D) The corresponding profile of the relative thickness (t/$\lambda$) calculated with the log-ratio method. $\lambda$ denotes the mean free path of ~100 nm. The thickness of the sample at the interface therefore equals ~ 30 nm.

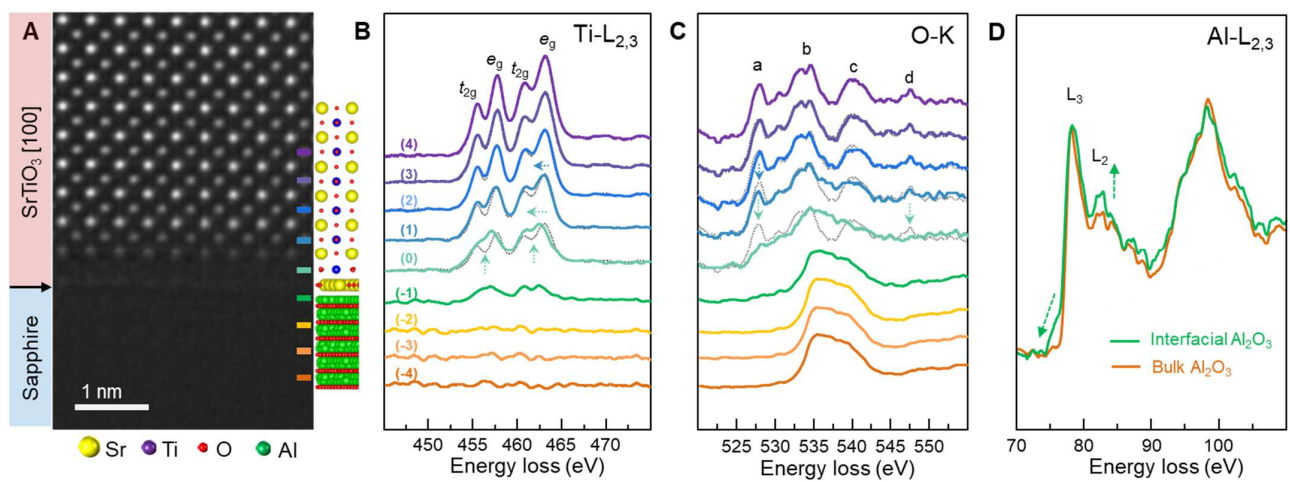

**Fig. S14.** Atomic-scale measurement of the electronic structure of a SrTiO$_3$/sapphire heterojunction across the interface. (A) Representative HAADF-STEM image of the SrTiO$_3$-sapphire interface with a schematic illustration of the derived sample structure at the right. The red arrow marks the interface. Colored lines indicate the atomic layers at which the EELS spectra were extracted. (B) EELS spectra of Ti-L$_{2,3}$ edges at different atomic planes across the interface, showing the clear red-shift of Ti-L$_{2,3}$ peaks and filling between $e_g$ and $t_{2g}$ peaks at the first SrTiO$_3$ layer next to the interface, as indicated by lateral and vertical arrows. (C) EELS spectra of O-K edge compared with the O-K edge spectrum from bulk SrTiO$_3$ (black dotted lines), showing a gradual drop of the intensity of peak (a) while approaching the interface and disappearance of peak (d) at the first SrTiO$_3$ layer next to the interface. (D) EELS spectra of Al-L$_{2,3}$ in the bulk Al$_2$O$_3$ and near the interface, revealing changed local Al-O configurations and the electronic structure close to the interface, indicated by the reduced threshold energy of the L$_3$ edge and the L$_3$/L$_2$ ratio.

**Supplementary Note 2:** The observed modulations of the electronic structure across the interface were studied with high spatial and high energy resolution STEM-EELS. Atomically resolved STEM-EELS spectra with an energy resolution of 0.6 eV were extracted to probe electronic states across the interface (Figs. S14A, S13). Background-subtracted EELS spectra of Ti-L$_{2,3}$ and O-K edges at each atomic layer are plotted in these figures, with reference spectra from bulk SrTiO$_3$ given as a black dotted line (Figs. S14B-C). Remarkably, the fine structure of the Ti-L$_{2,3}$ edges changes apparently across the interface, with peaks of the Ti-L$_{2,3}$ edges shifting towards lower energy. Approaching the interface from the SrTiO$_3$ side, the enhanced filling is observed between the $e_g$ and $t_{2g}$ peaks, particularly at the first SrTiO$_3$ layer next to the interface. (Fig. S14B). Meanwhile, the intensity of the pre-peak (a) of the O-K edge, which represents the filling of Ti 3d orbitals (*49*), shows a clear drop at the interface (Fig. S14C). These fine structure variations jointly demonstrate that there are noticeable reductions from $Ti^{4+}$ to $Ti^{3+}$ and also significant changes in the orbital bonding environment shift at the first $SrTiO_3$ layer next to the interface (50). The valence state of Ti in the first $SrTiO_3$ layer next to the interface. is measured to equal 3.60 $\pm$ 0.15. Therefore, electronic charge accumulates, possibly caused by oxygen defects. Beam-broadening effects give rise to the very weak $Ti-L_{2,3}$ edge signals across the interface, while the EELS map shows negligible Ti signal in this layer (Fig. S14B). Furthermore, peak (d), assigned to the transition from O 1$s$ to unoccupied O 2$p$ which is hybridized with Ti 4sp orbitals, disappears at the first $SrTiO_3$ layer next to the interface, revealing a deviation of the Ti-O coordinate environment from the TiO octahedral coordination (51). $Al-L_{2,3}$ edges correspond to the electron excitations from the Al-2p to Al-3s orbitals (52), fingerprinting the Al coordination environment and the unoccupied conduction band minimum. The threshold energy of $L_3$ is decreased by ~1.0 eV, and the $L_3/L_2$ ratio is substantially reduced at the interfacial Al layer compared to the bulk (Fig. S14D), reflecting an altered Al coordination symmetry and a band gap reduction (53). This band structure change is a candidate to account for the electronic charge accumulated at the interface.

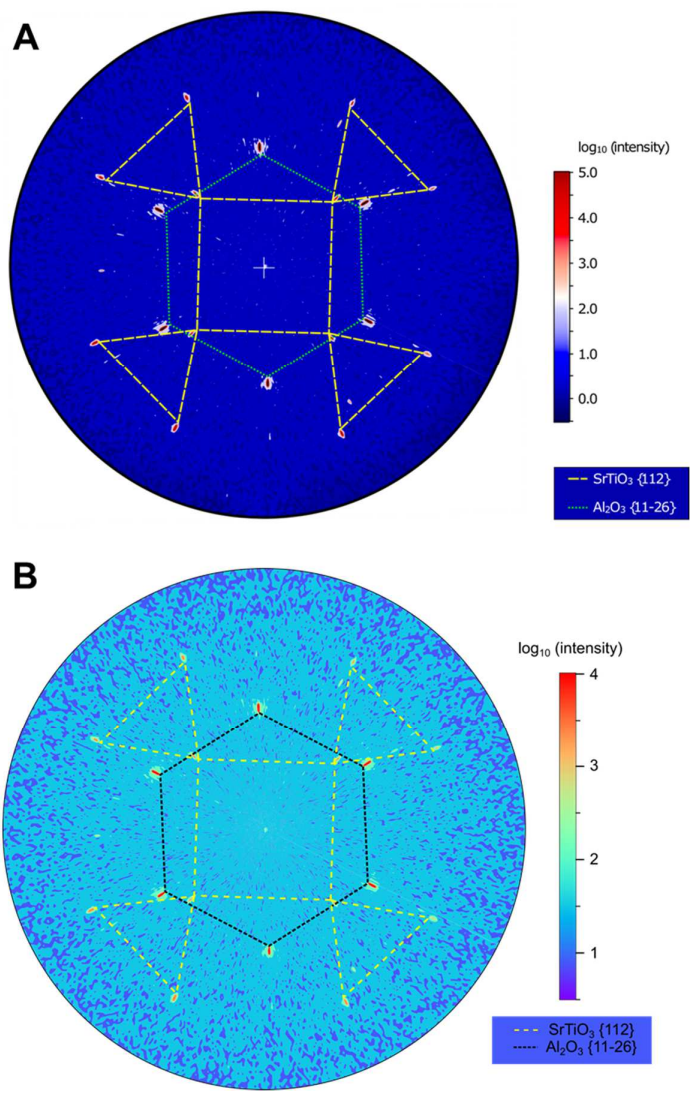

**Fig. S15.** (A) Fully annotated pole figure of the data shown in Fig. 4 of the main text. The complete family of SrTiO$_3$ {112} planes is identified and marked. The 12 peaks from the {112} family are well visible. Their positions make the 4-fold symmetry of the SrTiO$_3$ apparent. (B) The data shown here is the same as that shown in Fig. 4 of the main text. The color scale has been chosen to make the background more apparent.

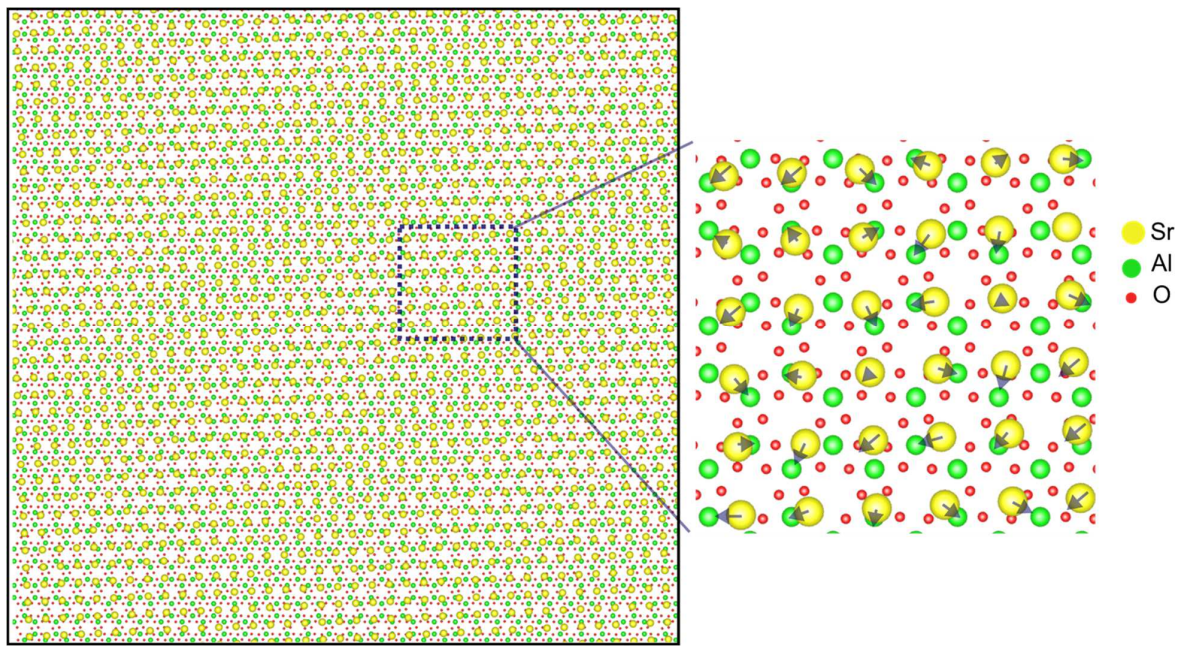

**Fig. S16.** The moiré pattern formed by the placement of the (001) Sr-O lattice of $SrTiO_3$ atop a singly terminated sapphire lattice for an in-plane twist angle of 3.5° between the $SrTiO_3$ [100] and the $Al_2O_3$ [1-100] axes. The arrows shown in the zoom-in on the right depict the local shifts that the Sr ions need to undertake in order to move closer to their neighboring Al ions. On this local scale, no spatial order is apparent in the pattern of these shifts, this may even seem random. However, on the scale of the moiré pattern resulting from the $SrTiO_3/Al_2O_3$ superposition, the shifts are governed by the quasi-periodicity of the moiré pattern.

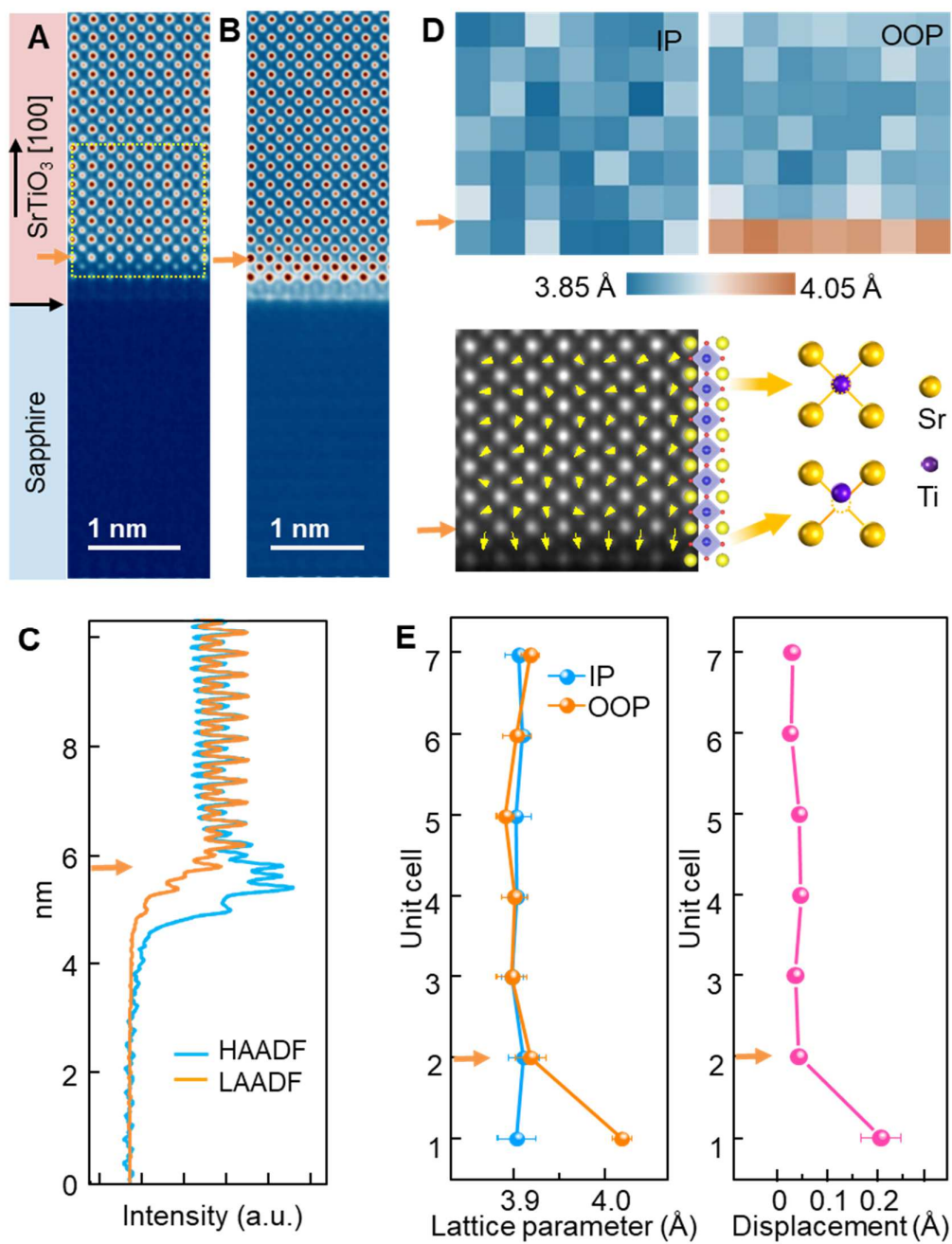

**Fig. S17**. Local structural distortion and polarity at the interface. (A)-(C) Representative atomic-resolution images of the interface viewed in HAADF (A) and LAADF (B) modes along the SrTiO₃ [100] direction, showing an apparent enhancement of the LAADF signal intensity (C). The black arrow denotes the out-of-plane direction. The vertical dark arrow marks the out-of-plane (OOP) direction. The orange arrow marks the second SrTiO₃ layer next to the interface. (D) Maps of the measured lattice constant taken along the in-plane (IP) (D, top-left panel) and OOP directions (D, top-right panel) and atomic displacement map (D,

bottom panel) of the square regions marked yellow in (A). (E) Laterally averaged profiles of lattice parameters and displacements measured along the out-of-plane direction, showing an apparent lattice expansion and polarization in the second SrTiO$_3$ layer next to the interface.

**Supplementary Note 3:** To further explore the interface microstructure, HAADF and low-angle ADF (LAADF) images of the same region were taken for comparison (Figs. S17A, B). The LAADF signal intensity of the first two SrTiO$_3$ layers next to the interface is enhanced, indicating the presence of a local structural disorder or strain that causes the dechanneling (Fig. S17C). Next, a quantitative analysis was performed to analyze the subtle but well notable structural distortion of the SrTiO$_3$ lattice adjacent to the first SrTiO$_3$ layer next to the interface. 2D maps of IP and OOP lattice parameters of a 7 uc x 7 uc region (the square region of Fig. S17A) show that the OOP lattice parameter is significantly enhanced in the SrTiO$_3$ layer adjacent to the first SrTiO$_3$ layer next to the interface. (Fig. S17D, top-right panel), while the IP lattice remains almost unchanged throughout the entire region (Fig. S17D, top-left panel). These data reveal an OOP tensile strain of 2.8% and an enhanced tetragonality of 1.03 of the SrTiO$_3$ layer adjacent to the first SrTiO$_3$ layer next to the interface., giving rise to a polar SrTiO$_3$ structure (Fig. S17D, bottom panel) characterized by a displacement between the Ti atomic column and the center of 4 neighboring Sr atomic columns by 20$\pm$4 pm (*54*) (Fig. S17E). Thus the local polarization is derived to equal 50$\pm$10 μC cm$^{-2}$ which requires a local screening charge density of 0.47$\pm$0.09 e$^-$/uc in the first SrTiO$_3$ layer next to the interface, in good agreement with the electronic charge observed by the EELS fine-structure analysis (Figs. S14B, C).

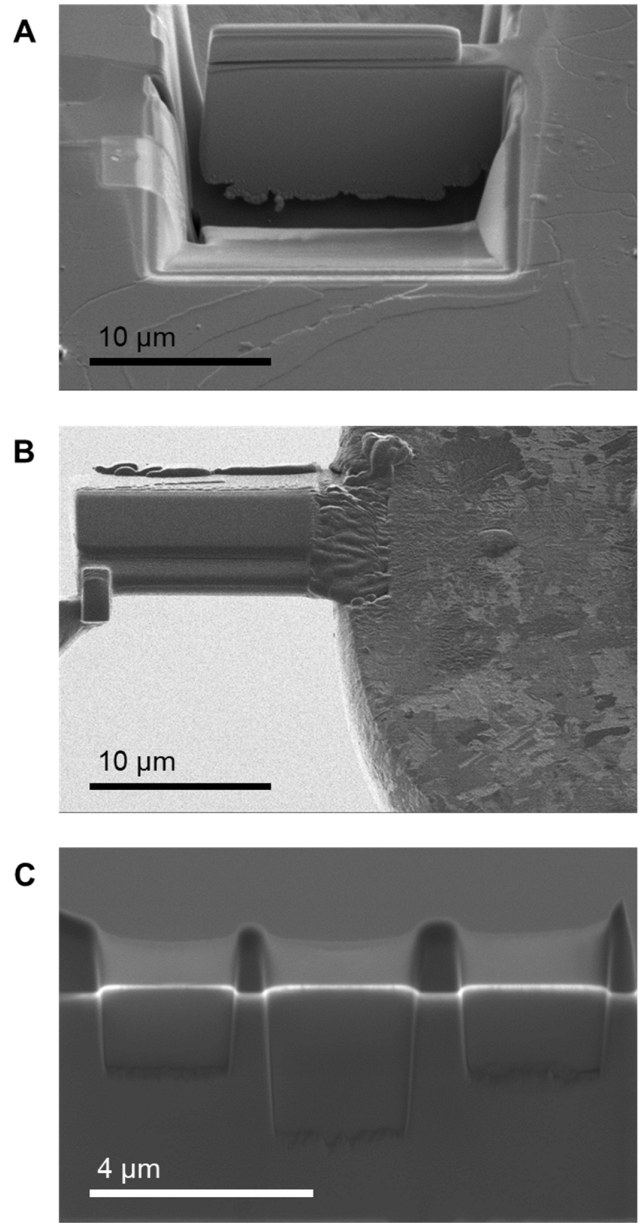

**Fig. S18.** Illustration of the TEM sample preparation using a focused ion beam. Two trenches were dug with a Ga⁺ ion beam to leave a thin section of material isolated at the center, which was then separated from the bulk with a U-shaped undercut (A). The lamella was lifted out with a micromanipulator and attached to a half-moon-shaped copper (Cu) grid (B). The FIB lamella was finally thinned by the Ga⁺ ion to a thickness below 50 nm for the region of interest (C).

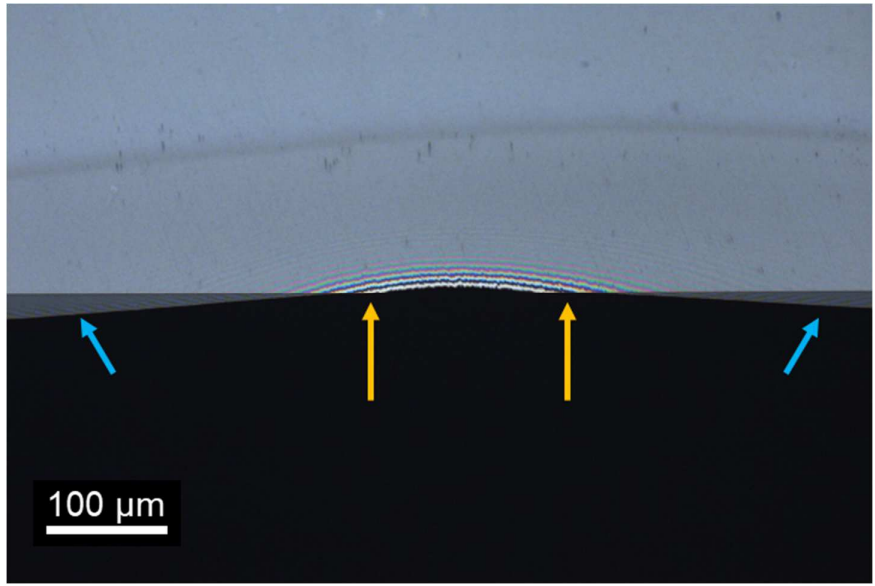

**Fig. S19.** Optical microscopy image of the thin region of the TEM specimen prepared by tripod polishing with subsequent Ar$^+$-ion milling. The central regions, showing optical interference fringes, are electron-transparent. The yellow arrows mark the areas of interest; the blue arrows mark the glass protection layer.

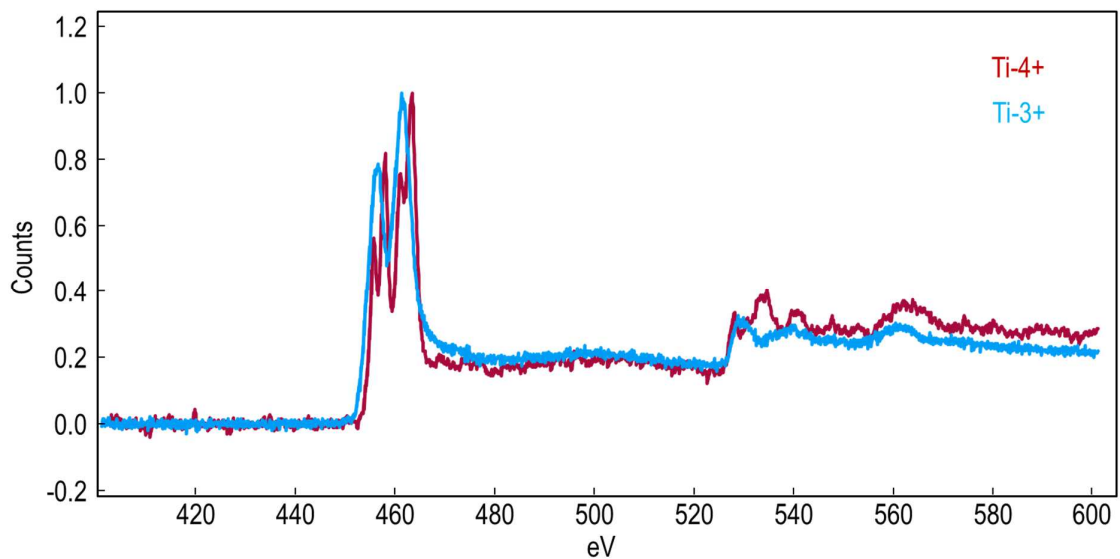

**Fig. S20**. EELS spectra of Ti-L$_{2,3}$ and O-K edges taken for calibration using standard SrTiO$_3$ (Ti-4+) and Li-TiO$_2$ (Ti-3+), respectively. The valence state of Ti was determined by multiple linear least-squares (MLLS) fitting the Ti-L$_{2,3}$ edge with the reference spectra of Ti$^{4+}$ and Ti$^{3+}$.

**References**

1. H. Y. Hwang *et al.*, Emergent phenomena at oxide interfaces. *Nature materials* **11**, 103-113 (2012).

2. J. Mannhart, D. Schlom, Oxide interfaces—an opportunity for electronics. *Science* **327**, 1607-1611 (2010).

3. J. Chakhalian *et al.*, Orbital reconstruction and covalent bonding at an oxide interface. *Science* **318**, 1114-1117 (2007).

4. B. H. Goodge *et al.*, Resolving the polar interface of infinite-layer nickelate thin films. *Nature materials* **22**, 466-473 (2023).

5. K. v. Klitzing, G. Dorda, M. Pepper, New Method for High-Accuracy Determination of the Fine-Structure Constant Based on Quantized Hall Resistance. *Physical review letters* **45**, 494-497 (1980).

6. A. Tsukazaki *et al.*, Quantum Hall effect in polar oxide heterostructures. *Science* **315**, 1388-1391 (2007).

7. G. Binasch, P. Grünberg, F. Saurenbach, W. Zinn, Enhanced magnetoresistance in layered magnetic structures with antiferromagnetic interlayer exchange. *Physical Review B* **39**, 4828-4830 (1989).